\documentclass[
amsmath, 
amssymb,
superscriptaddress,
twocolumn,
noshowpacs,
prl,
aps,
10pt,
longbibliography
]{revtex4-2}

\usepackage[T1]{fontenc}
\usepackage{enumerate}

\usepackage{siunitx}
\DeclareSIUnit\angstrom{\text {Å}}

\usepackage{graphicx}
\graphicspath{{Figures/}}

\usepackage{dcolumn}
\usepackage{braket}
\usepackage{bm}
\usepackage{color}

\newcommand{\beginsupplement}{
\setcounter{table}{0}
\renewcommand{\thetable}{S\arabic{table}}%
\setcounter{figure}{0}
\renewcommand{\thefigure}{S\arabic{figure}}%
}
\newcolumntype{C}{ >{\centering\arraybackslash} m{2cm} }

\newcommand{\Eu}{Eu\textsubscript{5}Sn\textsubscript{2}As\textsubscript{6}}

\begin{document}
\title{Colossal Magnetoresistance and Phonon Driven Exchange Dynamics in \Eu{}}

\newcommand{\UCB}{%
    Department of Physics, University of California, Berkeley, California 94720, USA
}
\newcommand{\LBNL}{%
    Materials Science Division, Lawrence Berkeley National Laboratory, Berkeley, California 94720, USA
}
\newcommand{\LANL}{%
    National High Magnetic Field Laboratory, Los Alamos National Laboratory, Los Alamos, New Mexico 87545, USA
}

\newcommand{\LBNLmol}{%
    Molecular Foundry, Lawrence Berkeley National Laboratory, Berkeley, California 94720, USA
}

\author{Luke \surname{Pritchard Cairns}}
\altaffiliation[Present Address: ]{Department of Physics, University of Tokyo, 7-3-1 Hongo, Bunkyo-ku, Tokyo 113-0033, Japan}
\affiliation{\UCB{}}

\author{Kohtaro \surname{Yamakawa}}
\affiliation{\UCB{}}
\affiliation{\LBNL{}}

\author{Shengzhi \surname{Zhang}}
\affiliation{\LANL{}}

\author{Youzhe \surname{Chen}}
\affiliation{\UCB{}}
\affiliation{\LBNL{}}

\author{Bernard \surname{Field}}
\affiliation{\LBNL{}}
\affiliation{\LBNLmol{}}

\author{Rainer \surname{Reczek}}
\affiliation{\UCB{}}

\author{Ryan P. \surname{Day}}
\affiliation{\UCB{}}

\author{Joel E. \surname{Moore}}
\affiliation{\UCB{}}
\affiliation{\LBNL{}}

\author{Marcelo \surname{Jaime}}
\altaffiliation[Present Address: ]{Physikalisch-Technische Bundesanstalt, Bundesalle 100, D-38116 Braunschweig, Germany}
\affiliation{\LANL{}}

\author{Sin\'ead M. \surname{Griffin}}
\affiliation{\LBNL{}}
\affiliation{\LBNLmol{}}

\author{Robert J. \surname{Birgeneau}}
\affiliation{\UCB{}}
\affiliation{\LBNL{}}

\author{James G. \surname{Analytis}}
\affiliation{\UCB{}}
\affiliation{\LBNL{}}
\affiliation{CIFAR Quantum Materials, CIFAR, Toronto, Canada}
\affiliation{Kavli Energy NanoScience Institute, Berkeley, CA, USA}

\date{\today}

\begin{abstract}
The emergence of colossal magnetoresistance in a new generation of Eu$^{2+}$-based antiferromagnets is intriguing given stark contrasts to the archetypal perovskite manganites and doped Eu-chalcogenides. In this study the thermal conductivity and magnetostriction of \Eu{}---one such representative---have been measured to better understand the role of the crystal lattice. Both properties are strongly field-dependent and mirror the magnetization, saturating once the Eu$^{2+}$ moments are polarized. The field-enhancement of the phonon-dominated thermal conductivity is interpreted through the lifting of a degeneracy of spin configurations, and the subsequent saturation due to quenched magnetostrain in high field. Comparison with spin-glass insulators suggests that this phenomenon is not a byproduct but rather the driver of electron delocalization due to the suppression of strong phonon scattering arising from exchange frustration.
\end{abstract}

\maketitle


The term `giant magnetoresistance' was coined to describe the large change in electrical resistance when polarizing certain antiferromagnetically coupled magnetic multilayers \cite{Baibich1988}. This phenomenon has revolutionized information technology and also given rise to the field of spintronics. Meanwhile colossal magnetoresistance (CMR)---so-called for the orders of magnitude enhancement of the effect---is yet to have any similar transformative impact, primarily due to the CMR existing at either too large an applied field or too low a temperature for practical applications. The two most heavily studied systems to exhibit CMR are intriguingly disparate in their respective properties, namely the manganite perovskite compounds \cite{Nagaev2001,Edwards2002,Tokura2006} and doped Eu(O,Se) \cite{Wachter1979,Mauger1986}. In the former, the interplay between double exchange---or enhanced electron hopping between ferromagnetically aligned neighbouring moments---and Jahn-Teller structural distortions is widely acknowledged to be pivotal to the CMR \cite{Millis1995,Visser1997,Cohn1997}. In the latter, ferromagnetic order instead appears to redshift the spin-split conduction band such that it overlaps with the defect levels \cite{Oliver1972,Arnold2008}. However, there is no consensus as to the precise mechanism in either case, something which has hindered efforts to engineer CMR into a practical phenomenon.

This study concerns \Eu{}, one member of an emerging group of next generation Eu$^{2+}$-based CMR compounds \cite{Chan1998_Eu14MnBi11,Sullow2000_EuB6,Devlin2018,Wang2021_EuCd2P2,Krebber2023_EuZn2P2} which struggle to be interpreted within either framework. Unlike the manganites there is no obvious analogue for double exchange, as the localised Eu$^{2+}$ pure spin moments couple very weakly to the surrounding crystal field and also contribute only indirectly to the charge conduction. Likewise, in contrast to doped Eu(O,Se) the material is a stoichiometric antiferromagnet, such that impurity levels and spin-split bands should not be relevant. Despite all this, \Eu{} exhibits magnetoresistance up to a factor of 6000, and the CMR persists over an comparatively broad temperature range---up to an order of magnitude larger than the ordering temperature, and to the lowest measured temperatures.

In this study, we attempt to understand the origin of CMR in \Eu{} and the broader class of compounds through an interpretation of thermal conductivity data. Despite the thermal conductivity being phonon dominated it is strongly enhanced with the application of field in the same temperature range which CMR manifests. Drawing analogy to amorphous solids we argue that, in zero-field, the proximity of \Eu{} to a spin glass transition leads to a collection of almost degenerate spin---and through magnetostrain, lattice--- configurations which can effectively scatter phonons. The application of field then polarises the spins, lifts this degeneracy and quenches the magnetostrain, resulting in a drastic reduction of phonon scattering. This picture is corroborated by magnetostriction and magnetisation data, which similarly saturate at high fields, and also classical spin modeling, which demonstrates that phonon modes can feasibly dynamically alter the magnetic ground state. A similar interpretation can also explain the large field-enhancement of the thermal conductivity observed in a number of spin glass insulators \cite{Arzoumanian1983,Lohneysen1987}. This phenomenon therefore occurs independently of electron localisation and might drive the CMR in \Eu{}.

\begin{figure}
    \centering
    \includegraphics[width=\columnwidth]{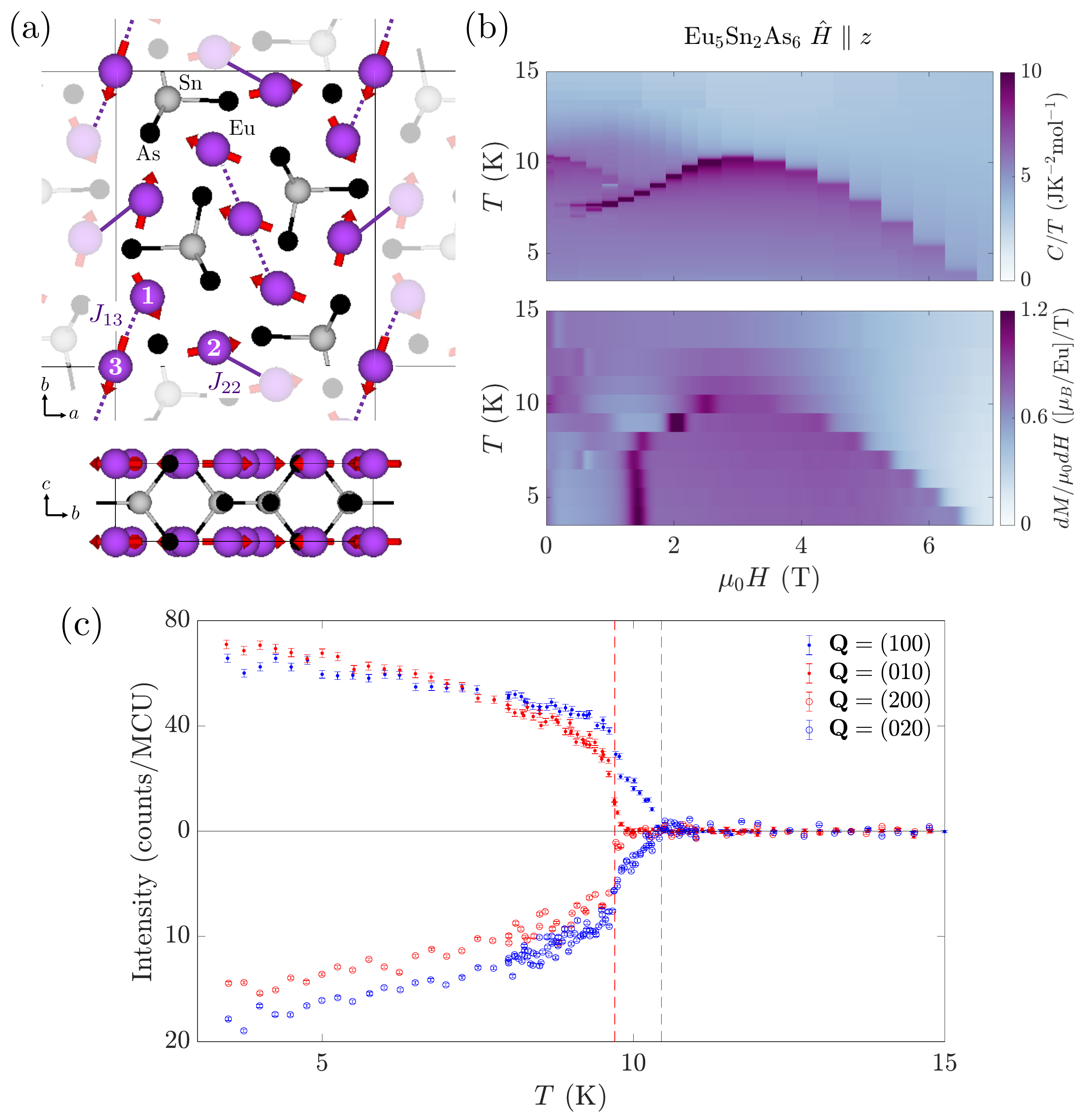}
    \caption{(a) Crystal structure of \Eu{}. 1, 2 and 3 label the three inequivalent Eu sites. $J_{22}$ and $J_{13}$ label the two nearest neighbour bonds, with separation between the Eu sites of {3.67876}~\SI{}{\angstrom} and {3.69799}~\SI{}{\angstrom} respectively. The ground state magnetic structure determined by neutron scattering is shown by the red arrows. (b) Contour plots showing the specific heat and field-derivative of the magnetisation for fields oriented $\hat{H} \parallel z$ ($\parallel c$ \cite{Day2025}). (c) Neutron diffraction results for \Eu{}, MCU is monitor counts collected for one second exposure of the neutron beam.  The transitions at $T_N$ and $T_{N'}$ are indicated by the vertical dashed lines.}
    \label{fig:Fig1}
\end{figure}

A thorough characterisation of \Eu{} was performed in a previous study \cite{Day2025}, which showed that a collection of almost-degenerate exchange pathways leads to a complex low-temperature magnetic phase diagram. For clarity we reproduce the crystal structure, and the low-temperature heat capacity and magnetisation for $\hat{H} \parallel z$ in Fig.~\ref{fig:Fig1}. In zero-field the material exhibits magnetic transitions at $T_{N'}=9.4$~K and $T_N=10.3$~K, and glassy dynamics---inferred through a slow relaxation of the magnetisation---in the narrow temperature range $T_{N'}<T<T_N$.  Despite these transitions being sharp and well-defined, they in fact release only a small fraction of the available entropy, and magnetic correlations persist to temperatures almost an order of magnitude larger than $T_N$. The CMR manifests up to a similar temperature, demonstrating that the electronic and magnetic subsystems are intimately linked.

For this study we performed inelastic neutron scattering in order to determine the magnetic structure of \Eu{}. As shown in Fig.~\ref{fig:Fig1}~(c), we observe magnetic Bragg peak intensities which emerge below $T_N$ at $\mathbf{Q} = (100)$ and $(020)$, and then below $T_{N'}$ at $\mathbf{Q} = (010)$ and $(200)$. As detailed in Appendix A, the zero-field, ground state magnetic structure is thus determined to be non-collinear antiferromagnetic with magnetic space group \textit{P2'/m'}. This has been superposed onto the crystal structure in Fig.~\ref{fig:Fig1}~(a). 

Within this magnetic structure, neighbouring planes along $z$ possess an identical configuration and are thus ferromagnetically correlated. This is in contrast to the closely related compounds Eu\textsubscript{5}In\textsubscript{2}Sb\textsubscript{6} and Eu\textsubscript{5}In\textsubscript{2}As\textsubscript{6} \cite{Balguri2025,Morano2024}, where instead nearest-neighbour moments along $z$ are antiparallel. This variability arises due to the simultaneous presence of FM indirect exchange and AFM superexchange interactions between Eu$^{2+}$ moments \cite{Kasuya1970}, which vary with the Eu-Eu and Eu-(As,Sb) separation respectively. The competition is exemplified beautifully by the aforementioned Eu-based chalcogenides---as the inter-atomic separation is increased the FM order is continuously weakened in Eu(O,S,Se), before culminating in AFM order in EuTe \cite{Wachter1979}. The same situation is replicated in this set of compounds, for which \Eu{} has the smallest inter-layer separation, and is the only compound to possess a ground state with parallel spins along the $z$-axis. This serves to demonstrate that the precise magnetic order is irrelevant to the CMR, which manifests similarly in all of the Eu-526 compounds.
    
\begin{figure}
    \centering
    \includegraphics[width=\columnwidth]{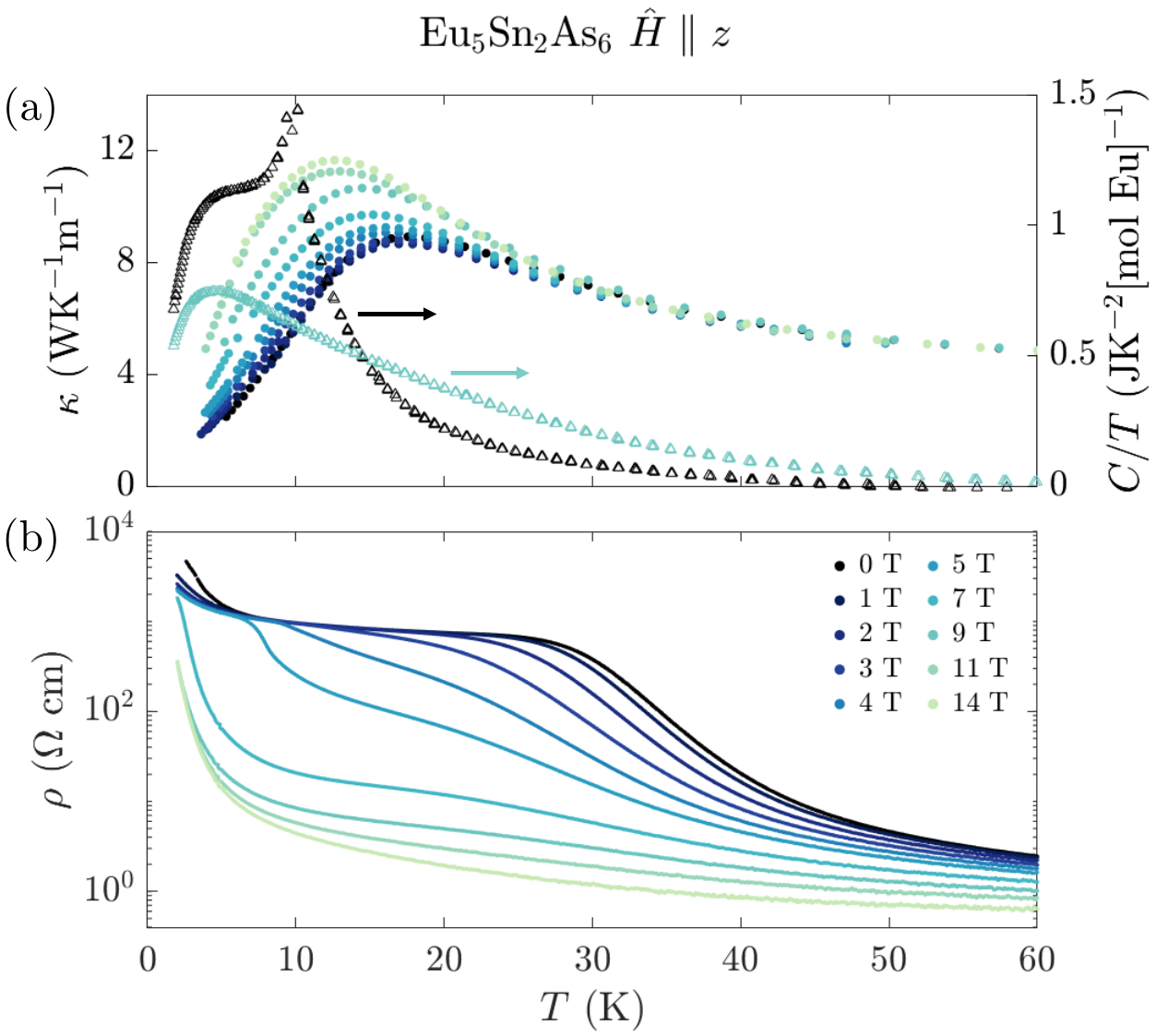}
    \caption{(a) Thermal conductivity of \Eu{} as a function of temperature for a range of fields applied $\parallel z$ (left axis, closed circles), and the specific heat at 0 and 9~T (right axis, open triangles), with the estimated phonon contribution subtracted (see SM).  (b) Electrical resistivity under the same conditions. In both cases the current is also applied $\parallel z$.}
    \label{fig:Fig2}
\end{figure}

Fig.~\ref{fig:Fig2}~(a) shows the temperature-dependent thermal conductivity for a range of fields oriented along the $z$-axis, alongside  specific heat and electrical resistivity data. In \Eu{}, the thermal conductivity is phonon-dominated across the full measured temperature and field range. This is inferred from (i) the Wiedemann-Franz law providing an approximate upper bound on the electronic heat current, which is orders of magnitude smaller than the measured thermal conductivity at all fields (see SM), and (ii) the thermal conductivity being continuous through the magnetic ordering transitions, implying the absence of a significant magnon contribution. Despite being phonon-dominated, the application of field acts to drastically enhance the thermal conductivity at the lowest temperatures. This field-enhancement survives up to temperatures $\sim40$~K---far larger than the ordering temperature. This also represents an approximate upper bound for the release of magnetic entropy and the occurrence of CMR. It is clear therefore that the lattice is also strongly tied to the magnetic and electronic subsystems. 

The observed field enhancement must be due to a reduction in phonon scattering via some electronic or magnetic channel. We can immediately discount the former as the drastic enhancement of the charge carrier mobility with the application of field would act to enhance the phonon scattering, counter to observations. For the latter, spin-phonon coupling in rare-earth systems typically occurs through single-ion striction \cite{Doerr2005}, or the shifting of the spin energy-levels through modulations of the surrounding crystal field. However, the ground state of Eu$^{2+}$ is a pure spin $J=S=\frac{7}{2}$ multiplet and so this mechanism can provide only a very weak coupling. Similarly, resonant phonon scattering from Zeeman split levels struggles to explain the data, as this would lead to a enhanced phonon scattering with increasing field. The strongest scattering mechanism that remains is exchange striction, or the modulation of the exchange strength upon altering the zeparation between interacting spins \cite{Luthi2007}. As demonstrated by the comparison between the Eu-526 compounds above, the exchange between Eu$^{2+}$ moments depends sensitively on separation and it is thus reasonable that exchange striction should be important. This is similarly the case for a variety of Gd$^{3+}$ pure-spin materials \cite{Rotter2001,Massalami2003,Correa2012,Lee2013}.

In the exchange striction model \cite{Doerr2005}, the magnetostrictive strain tensor is defined as
\begin{equation}
    \epsilon^\alpha = \frac{1}{2} \sum_{\beta,ij} s^{\alpha \beta} \langle\mathbf{J}_i \mathcal{J}_{(\beta)}(ij)\mathbf{J}_j\rangle_{T,\mathbf{H}},
\end{equation}
where $s^{\alpha \beta}$ is the elastic compliance, $\langle \rangle_{T,\mathbf{H}}$ is the thermal expectation of the static correlation and
\begin{equation}
    \mathcal{J}_{(\beta)}(ij) = \bigg[\frac{\partial \mathcal{J}(ij,\epsilon)}{\partial \epsilon^\beta}  \bigg]
\end{equation}
is the two-ion interaction strength, assuming an isotropic exchange. In the absence of spin-orbit coupling---which would lead to single-ion contributions---the magnetostrain will depend only on the relative orientation of interacting moments. 

\begin{figure}
    \centering
    \includegraphics[width=0.9\columnwidth]{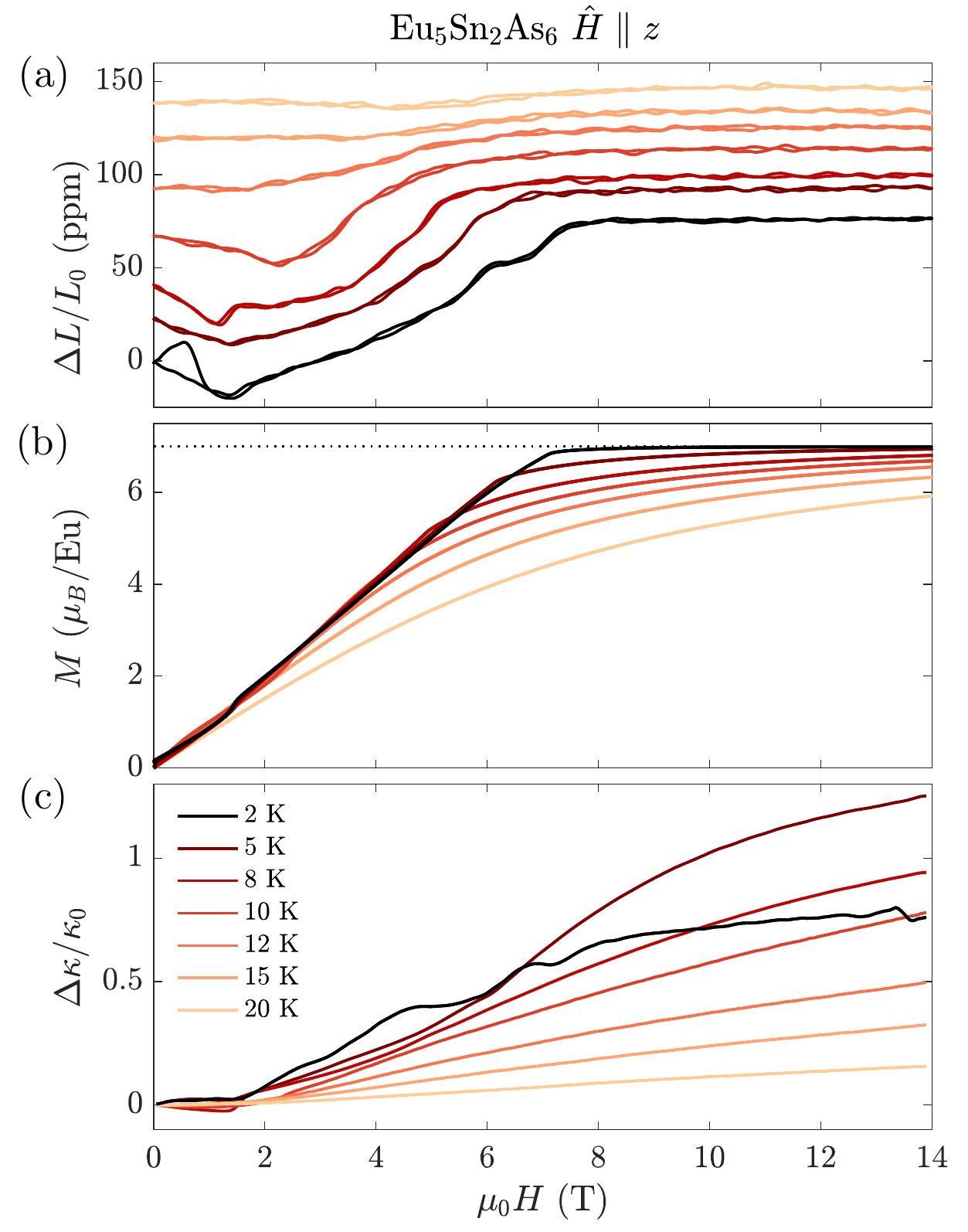}
    \caption{(a) Magnetostriction, (b) magnetisation, and (c) thermal conductivity of \Eu{} as a function of applied field $\hat{H} \parallel z$, at a range of temperatures. $\kappa_0$ and $L_0$ refer to the values at zero-field. The magnetostriction data has been offset for clarity. The thermal gradient is also applied $\parallel z$. In (b), the dashed line is at $7\mu_B$/Eu, the expected saturation moment of a free Eu$^{2+}$ ion.}
    \label{fig:Fig3}
\end{figure}

The impact of magnetostrain is clearly demonstrated by the field-dependent magnetostriction data for \Eu{}, shown in Fig.~\ref{fig:Fig3}~(a). The zero-field magnetic state is a frustrated configuration of spins with essentially no satisfied bonds. Upon the application of field the moments are forced to rotate, which alters the relative orientations and therefore the magnetostrain. The magnetic phase diagram (Fig.~\ref{fig:Fig1}~(b)) is reflected in the various sharp features and character of the magnetostriction in each region. At the largest fields and the lowest temperatures the magnetisation saturates (Fig.~\ref{fig:Fig3}~(b)) as the moments are fully polarized; the magnetostrain similarly saturates as the respective orientations of the moments become fixed.

The field-dependence of the thermal conductivity---shown in Fig.~\ref{fig:Fig3}~(c)---can also be explained phenomenologically through magnetostrain. In zero-field \Eu{} teeters on the brink of a transition to a spin glass, with a variety of almost degenerate magnetic configurations \cite{Day2025}. Crucially, magnetostrain implies there is equivalently a variety of almost degenerate lattice configurations, which can be cycled between via classical thermally activated barrier hopping and the inelastic scattering of phonons \cite{Prejean1980}. Due to there not being a single energy scale, this provides an effective scattering mechanism for a large portion of the phonon spectrum and the thermal conductivity is consequently low. With the application of field certain magnetic configurations become energetically favourable before, at the largest fields, a gap opens to the fully polarized state. Once this gap grows larger than the highest energy phonon mode there can no longer be scattering via this mechanism and the thermal conductivity saturates at its largest value. 

This framework borrows heavily from the theory of amorphous solids. Specifically, the excessively low thermal conductivity in amorphous solids at low temperatures has been interpreted as due to transitions between almost degenerate structural configurations \cite{Anderson1972,Phillips1972,Muller2019}. These effectively present a constant density of states for the scattering of acoustic phonons, which massively suppresses the thermal conductivity \cite{Pohl2002}. Unlike the amorphous solids however, for \Eu{} the application of field acts to lift this degeneracy and the thermal conductivity is thus enhanced. 

An enhancement of the thermal conductivity with applied field has also been observed in a variety of spin-glass materials \cite{Arzoumanian1983,Wassermann1984,Ma2018}, and in the case of Eu$_{0.4}$Sr$_{0.6}$S it grows to a factor of 5 times larger. This material is particularly illuminating as the compositional series Eu$_{x}$Sr$_{1-x}$S also features ferromagnetism ($x \gtrsim 0.5$) and paramagnetism ($x \lesssim 0.1$), but an enhancement is only observed in the spin glass regime \cite{Lohneysen1987}. Importantly, the entire compositional series is insulating, which demonstrates that the enhanced phonon scattering is not some byproduct of electron localisation.

In order to demonstrate the feasibility of dynamical modulation of the magnetic ground state through phonon scattering, we have performed classical spin modeling using the SU(N)NY software \cite{Dahlbom2025}. Using a minimal model which mimics the well-established treatment of the Eu(O, S, Se, Te) compounds \cite{Kasuya1970} but also includes a small Dzyaloshinskii-Moriya interaction term ($\sim 0.01$~K) on symmetry-allowed bonds, we are able to reproduce the ground state magnetic structure of \Eu{}. Then, as shown in Fig.~\ref{fig:app_spin_model}, when distorting the lattice constants by by less than the estimated Debye-Waller factor ($\sim0.001$~\SI{}{\angstrom}$^2$ at 10~K \cite{Subhadra1992,Takahashi1999}), we find that the ground state spin configuration changes. This effect---or the capacity of acoustic phonons to dynamically alter the strength of magnetic exchange---has already been observed in a variety of systems \cite{Lovinger2020,Padmanabhan2022,Cong2022}.

Finally, in deciphering the origin of the CMR in \Eu~we can take hints from similar phenomena which occur in the manganites or in doped-EuO. As with the manganites, dynamical lattice \cite{Millis1995} and spin fluctuations \cite{Kogan1988} are relevant here and appear to dictate the low-temperature thermodynamic properties. However, density functional theory calculations (Appendix B) demonstrate that the conduction in \Eu{} primarily derives from the As $p$-orbitals rather than the Eu orbitals, such that there is no analogue for double exchange. In the case of EuO, CMR only manifests upon doping (in EuO$_{1-x}$ or Eu$_{1-x}$Gd$_x$O for example \cite{Oliver1972,Mauger1986}), and is thus strongly impacted by the degree of disorder. Disorder also appears to be integral to the CMR in \Eu{}, but instead manifests through magnetostrain rather than random site mixing. This might also be linked to the idea of magnetic polarons, wherein charge carriers will polarize---and then become localized within---a small cluster of local moments; a concept which has been put forward to explain the CMR in all of these materials \cite{Batista2000,Torrance1972,Rosa2020}. Recent studies have demonstrated that an appropriate level of structural disorder can facilitate magnetic polaron formation \cite{Mondal2025}---perhaps this is the role of magnetostrain in \Eu{}.

One important aspect in which \Eu{} is different to the majority of other compounds to exhibit CMR is that the phenomenon persists to the lowest measured temperature, rather than existing only within a close proximity to the ordering temperature \cite{Schiffer1995,Oliver1972,Wang2021_EuCd2P2}. Similar to the broadband phonon scattering observed in amorphous solids, this implies an effectively constant density of states for electron localisation over a wide energy range. The glassy nature of \Eu{} and resulting collection of almost degenerate spin (and therefore lattice) configurations in zero field provide exactly this. This also implies that---much like an applied field---lattice deformations should be capable of lifting this degeneracy and increasing the conductivity, suggesting that this phenomenon might find utility in strain or pressure sensing. Indeed, colossal piezoresistance has already been demonstrated in the sister compound Eu$_5$In$_2$Sb$_6$ \cite{Ghosh2022}. 

In this study we have demonstrated that the spin, charge and lattice subsystems in \Eu{} are intimately linked and that exchange striction is the dominant mechanism for spin-phonon coupling, which drives the low-temperature thermodynamic properties. The low thermal and electrical conductivities (and anomalously broad heat capacity) in zero-field can be interpreted through a collection of almost degenerate spin---or equivalently lattice---configurations, which offer an effectively constant density of states for the scattering of thermal and charge carriers. The role of the lattice in the emergence of CMR is evident here, and appears similarly relevant to a variety of disparate compounds. 


\begin{acknowledgments}
\section{ACKNOWLEDGEMENTS}
This work was primarily funded by the U.S. DOE, Office of Science, Office of Basic Energy Sciences, Materials Sciences and Engineering Division under Contract No. DE-AC02-05CH11231 (Quantum Materials Program KC2202). The work performed at the Molecular Foundry (BF, SG) was supported under the same contract within the Theory of Materials program. Computational resources were provided by the National Energy Research Scientific Computing Center and the Molecular Foundry, DOE Office of Science User Facilities. Work by KY was performed with support through a MURI project supported by the Air Force Office of Scientific Research (AFSOR) under grant number FA9550-22-1-0270. A portion of this work was performed at the National High Magnetic Field Laboratory,  supported by the NSF Cooperative Agreement No. DMR-1644779, the US DOE and the State of Florida.

\end{acknowledgments}


\appendix

\section{Appendix A: Refinement of the Magnetic Structure} \label{AppendixA_neutron}

\begin{figure}[b]
    \centering
    \includegraphics[width=1\columnwidth]{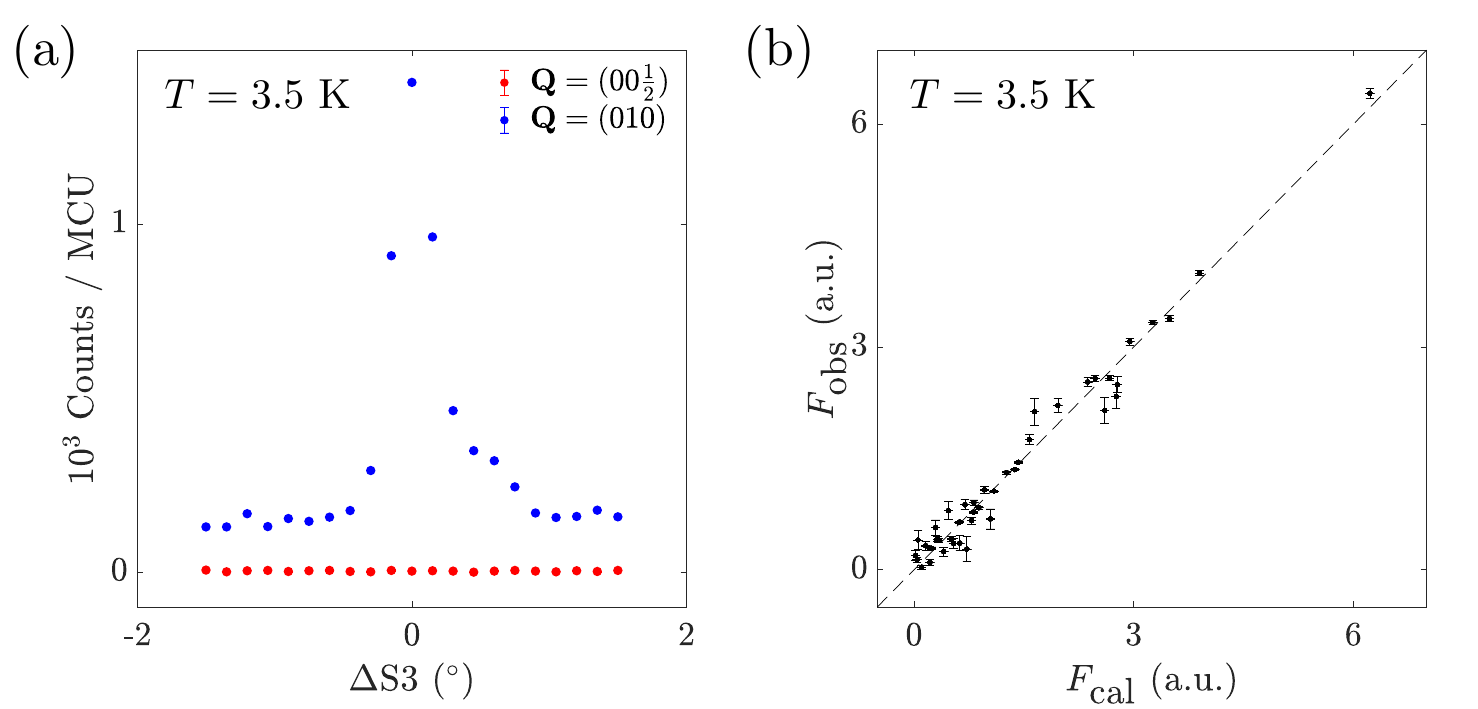}
    \caption{(a) Rocking scans performed around \( \mathbf{Q} = (00\frac{1}{2}) \) and \( \mathbf{Q} = (010) \) (b) Observed versus calculated magnetic structure factors ($F_{\mathrm{obs}}^{2}$ vs.\ $F_{\mathrm{cal}}^{2}$) for the refined model.}
    \label{fig:FigApp_NS}
\end{figure}

Given that the order parameter evolving continuously suggests a second-order phase transition, we model the magnetic structure of \Eu{} using irreducible representation (IR) analysis based on the paramagnetic space group \textit{Pbam}, as determined by room-temperature powder X-ray diffraction. As shown in Fig.~\ref{fig:FigApp_NS}~(a), rocking scans at 3.5~K demonstrate that magnetic scattering is observed only at the positions of integer numbers in the reciprocal lattice unit. The magnetic propagation vector is thus identified as \(\mathbf{k} = 0\). The relevant IRs of the little group \( G(\mathbf{k}=0) \) are \( \Gamma_{1},~\Gamma_{3},~\Gamma_{5}, \) and \( \Gamma_{7} \), with their associated basis vectors (BV) provided in Table~\ref{tab: BVs}. The selection rules for the magnetic Bragg peak positions corresponding to each IR/BV are summarized in Table~\ref{tab: selection}, along with the experimental observations.

For the second-order phase transition from the paramagnetic phase at \( T_{N} \), Landau theory requires that the magnetic structure be described by a single IR. Among the possible IRs, only \( \Gamma_{7} \) is consistent with the experimental observations. In this configuration (\textit{Pb'am'}), the \( y \)-components of spins on the same Wyckoff site are aligned antiferromagnetically, whereas the \( x \)-components are aligned ferromagnetically.

For the additional IRs involved below $T_{N'}$, either a single IR ($\Gamma_{5}$) or a combination of two IRs ($\Gamma_{1} + \Gamma_{3}$) could in-principle be used to describe the magnetic structure according to the selection rules. To determine which IR is realized, the symmetry elements of the magnetic structure for each IR are listed in Table~\ref{tab:sym}. Our analysis shows that the sequences $\Gamma_{7} \rightarrow \Gamma_{1} + \Gamma_{7} \rightarrow \Gamma_{1} + \Gamma_{3} + \Gamma_{7}$ and $\Gamma_{7} \rightarrow \Gamma_{3} + \Gamma_{7} \rightarrow \Gamma_{1} + \Gamma_{3} + \Gamma_{7}$ would break the symmetry twice, which is inconsistent with a single second-order phase transition at $T_{N'}$. This leaves $\Gamma_{5} + \Gamma_{7}$ as the only viable option to describe the low-temperature magnetic structure.

The magnetic structure of \Eu{} was refined using a model described by $\Gamma_{5}+\Gamma_{7}$ against the elastic neutron scattering data, revealing a non-collinear antiferromagnet with a magnetic space group identified as \textit{P2'/m'}, as shown in Fig.~\ref{fig:Fig1}~(a). The observed structure is compared with the calculated structure factors in Fig.~\ref{fig:FigApp_NS}~(b) and found to be consistent within the experimental error bars.

\section{Appendix B: Electronic Structure Calculations} \label{AppendixB_DFT}

\begin{figure}
    \centering
    \includegraphics[width=0.8\columnwidth]{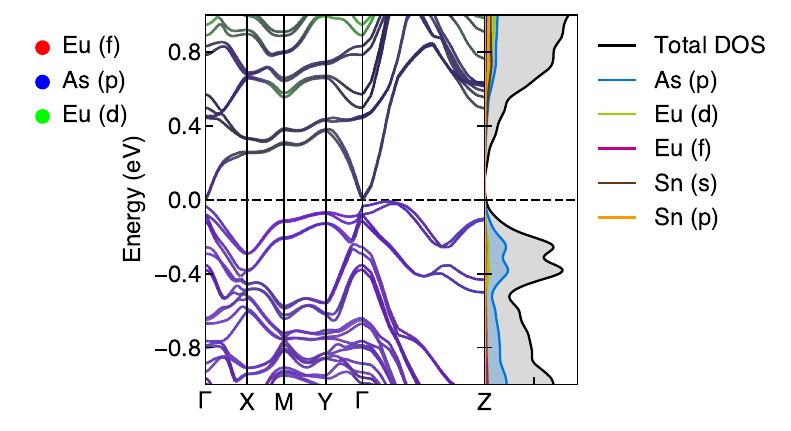}
    \caption{Electronic band structure and density of states, including the contribution from each orbital. Calculated at $U = 10$~eV. Plotted with sumo \cite{Ganose2018}.}
    \label{fig:app_DFT}
\end{figure}

Electronic structure calculations were carried out using density functional theory (DFT), details of the method are included in the SM. For DFT+U, a range of $U$ values were tested and compared against the case with frozen $f$-electrons. Comparing against electric transport measurements, the paramagnetic (no $f$-electron) case matches the Arrhenius-like behavior at higher temperatures (above the magnetic ordering temperature) with a gap of $\Delta = 38$~meV (see SM). This is similar to the experimental value of 35~meV \cite{Day2025}. This suggests that the $f$-band has negligible hybridization with the bands near the Fermi level, and we can take a large value of $U=10$~eV that pushes the $f$-band far from the Fermi level. Notably, these conditions also give the magnetic structure found via neutron scattering (Fig.~\ref{fig:Fig1}~(a)) as a converged solution. As shown in Fig.~\ref{fig:app_DFT}, the band projection gives the valence band to be mostly As-$5p$ in character, while the conduction band is mostly Eu-$5d$.  This supports the classification of \Eu{} as a Zintl phase material where the divalent europium cations 5[Eu]$^{2+}$ are ionically bonded by the polyanion [Sn$_2$As$_6$]$^{10-}$ group.  
\\

One of the unique aspects of \Eu{} compared to the other Eu-526 compounds is the appearance of a low temperature plateau in electrical resistivity. As speculated in a previous study \cite{Day2025}, one possible explanation for this shunting is through surface states of a topological insulator, such as in SmB$_6$ \cite{Neupane2013}. Indeed, by calculation of the magnetic topological symmetry indicators \cite{Xu2020,Elcoro2021}, \Eu{} was found to be a weak topological insulator, with topological indices $(Z_4;Z_{2,a}Z_{2,b}Z_{2,c})=(0;110)$. While this suggests the possibility of tuning topology by using magnetic field or doping to interpolate between magnetic structures, this is beyond the scope of this work.

\section{Appendix C: Classical Spin Modeling} \label{AppendixC_model}

We used the SU(N)NY software \cite{Dahlbom2025} to perform classical spin modeling, in order to investigate the feasibility of dynamical modulation of the magnetic ground state through phonon scattering. The magnetism in \Eu{} derives from the half-filled $4f$ orbitals of the Eu$^{2+}$ ions, and therefore comprises pure-spin $J=S=\frac{7}{2}$ moments with negligible overlap. This implies Heisenberg direct exchange should not be relevant. Similarly, due to the 35~meV band gap and resultant low charge carrier density \cite{Day2025}, RKKY interactions cannot mediate the magnetism. This is also the case for the EuX (X = O, S, Se, Te) compounds, for which there exists a well established---and successful---model \cite{Kasuya1970,Wachter1979}.

Within this model the dominant magnetic interactions occur through excitations to the unfilled Eu $5d$-shell, either from the Eu $4f$-shell---so-called indirect exchange, typically FM---or from the As $4p$-shell---$d$-$f$ exchange, typically AFM. The relevant parameters are therefore $J_{df}$ (the direct exchange between the $5d$ and $4f$ shells of the Eu ion), $U_{d[4f,4p]}$ (the energetic separation between the relevant $d$-orbital and either the Eu-$4f$ or As-$4p$ bands) and $t_{dn}$ (the overlap between the relevant $n$- and $d$-orbitals). $J_{df}$ should be approximately material-independent and has been measured to be $\sim100$~meV \cite{Pivetta2020}, and $U_{d[4f,4p]}$ can be estimated from the calculated band structure (Fig.~\ref{fig:app_DFT}). However, $t_{dn}$ requires knowledge of $d$-orbital splitting and orientation at each Eu site. This was determined for each site from a minimisation of the overlap between the lowest energy $d$-orbital and the surrounding crystal electric field from neighbouring As sites. The energy scale of $t_{dn}$ was set by assuming that the value should be similar for the Eu1-Eu3 bond in \Eu{} and the Eu-Eu bond in EuSe, which both occur between octahedrally coordinated Eu ions at almost identical separations, both surrounded by $4p$-orbitals.

From this, we can assign an interaction strength ${J_{[ij]} = J^{\text{FM}}_{[ij]} + J^{\text{AFM}}_{[ij]}}$ between Eu sites $i$ and $j$, then solve an effective Heisenberg-like Hamiltonian of the form ${\mathcal{H} = \sum_{ij} J_{ij} \mathbf{S}_i \cdot \mathbf{S}_j}$ to find the magnetic ground state. This treatment obviously respects the structural $Pbam$ space group whereas the magnetic ground state determined through neutron scattering splits each Wyckoff position into two inequivalent sites. Therefore, considering the canted nature of the true ground state, we also included a constant Dzyaloshinskii–Moriya (DM) term on symmetry allowed bonds. For ${J_{DM}=0.013}$~K (compared to, for example, ${J_{13} = 0.414}$~K) we were able to reproduce the magnetic structure shown in Fig.~\ref{fig:Fig1}~(a).

\begin{figure}
    \centering
    \includegraphics[width=1\columnwidth]{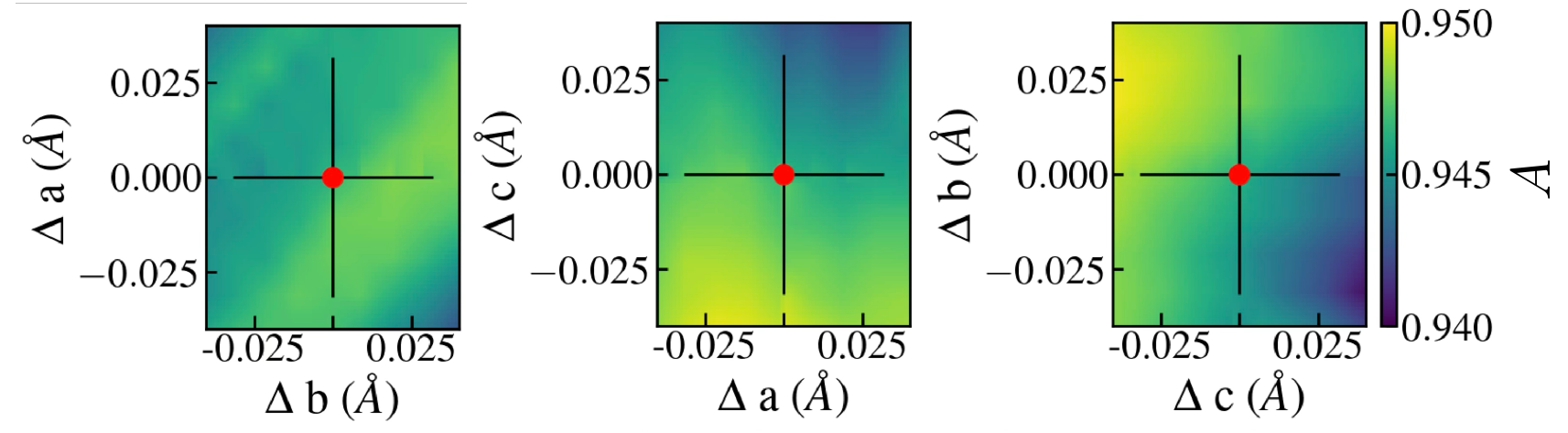}
    \caption{Variation of the spin alignment factor $A$ (described in the main text) with changing lattice parameters. In each case the red dot represents the lattice parameters of \Eu{} and the lines extend to the estimated maximum displacement due to phonon modes at 10~K.}
    \label{fig:app_spin_model}
\end{figure}

Finally, in order to determine whether a low-temperature phonon mode might feasibly alter the magnetic ground state, we repeated the same calculation for different distortions of the unit cell. The results are shown in Fig.~\ref{fig:app_spin_model}, where the relative change of the magnetic configuration is quantified using
\begin{equation}
    A = \frac{1}{N}\sum_i^N\frac{s_i \cdot S_i}{|s_i||S_i|}.
\end{equation}
$s_i$ and $S_i$ are the magnetic moments for the Eu site $i$ from the calculation and the neutron scattering result respectively, and $N=6$ is the number of independent sites. In changing the lattice parameters up to the anticipated phonon displacement at 10~K (estimated from the Debye-Waller factor in similar materials \cite{Subhadra1992,Takahashi1999}) we find that $A$ continuously changes. This is exactly the expectation for a variety of almost degenerate magnetic configurations. The change in alignment is obviously small in comparison to a spin flip, but it is important to recognize that the scattering will primarily be from clusters rather than single sites. This can be seen from the fact that the magnitude of the thermal conductivity far exceeds the theoretical minimum for single-site scattering, which can be calculated by assuming a random walk between anharmonic oscillators \cite{Cahill1992} (see SM). The wavelength of the dominant phonon mode at 10~K should therefore be $\sim10$~nm (from $\lambda \simeq \frac{h v_s}{3 k_B T}$ where $v_s$ is the sound velocity), which is the typical length scale expected for a cluster of correlated spins.

\bibliography{Eu526.bib}


\onecolumngrid
\newpage
\beginsupplement

\section{Supplementary Material}

\subsection{Experimental Methods}

\begin{figure}[h]
    \centering
    \includegraphics[width=0.6\columnwidth]{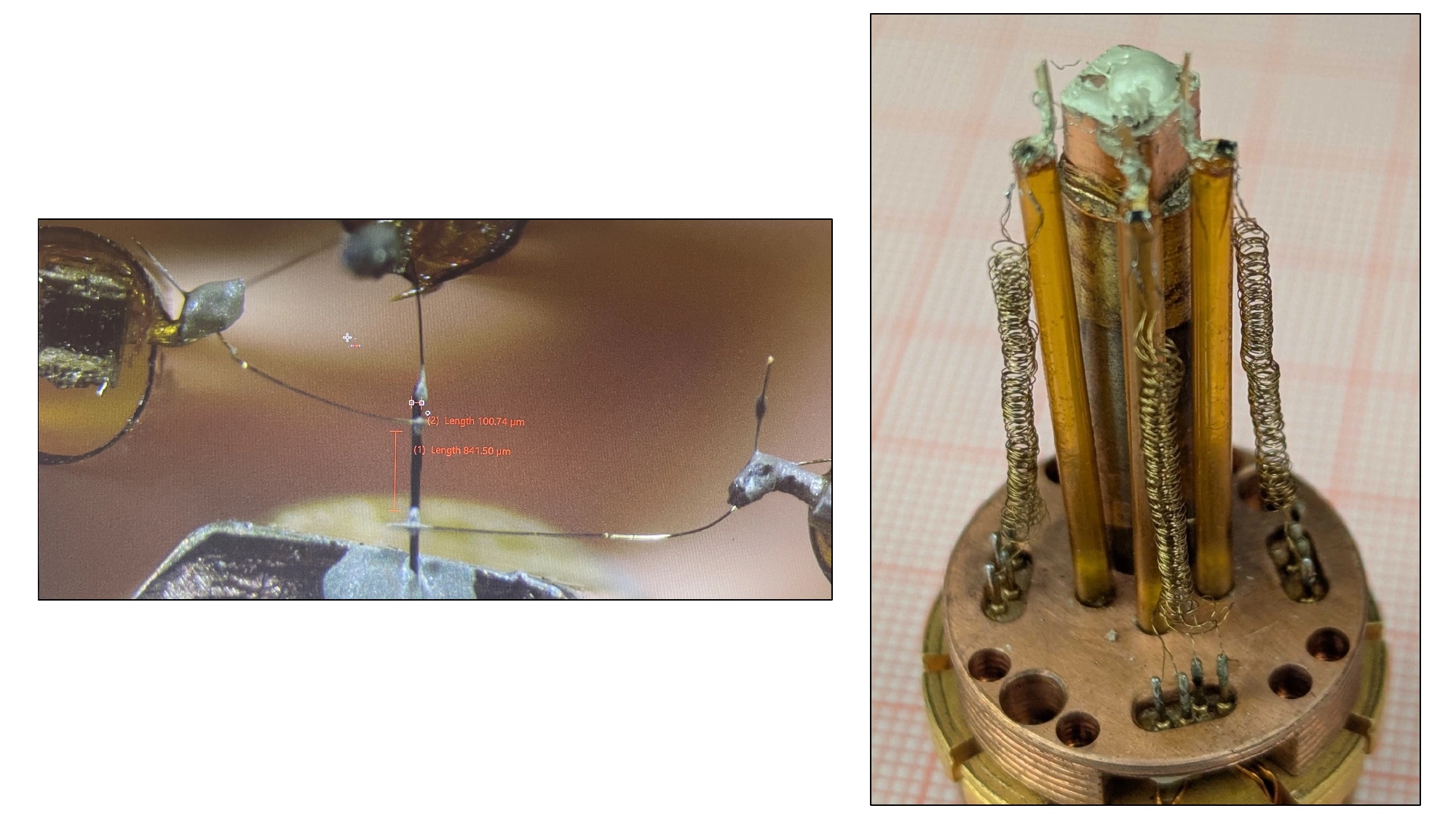}
    \caption{Thermal conductivity setup. The left figure shows a typical \Eu{} needle-like sample prepared for measurement. The right figure shows the full configuration, demonstrating the methods for thermal isolation of the heater and thermometers.}
    \label{fig:SM_setup}
\end{figure}

All measurements were performed on single crystals, using a Quantum Design (QD) PPMS Dynacool for magnetic field and temperature control. Magnetisation and heat capacity measurements were performed using the QD vibrating sample magnetometer and heat capacity options, with the latter employing both the relaxation-time and dual-slope methods. Dilatometry measurements were performed using a fiber Bragg grating method (described in \cite{Jaime2017}). Thermal conductivity measurements were performed using the home-built setup shown in Fig.~\ref{fig:SM_setup}, and measured using the steady-state one heater, two thermometers method with $\Delta T<0.03T$. The design is based on that from \cite{Kim2019}, but instead using resistive thermometers and polyimide support posts. Temperature-dependent calibration data was collected at discrete fields, then fit to the same order polynomial to give a calibration function for each field. For field sweep measurement, the coefficients of these calibration functions were interpolated between in order to generate a calibration at any intermediate field.
\\

Neutron diffraction measurements were performed using the Versatile Intense Triple-Axis Spectrometer (VERITAS, HB-1A) at Oak Ridge National Laboratory. Both the incident and final neutron energies were fixed at 14.5 meV. A collimation of 40$^{\prime}$–40$^{\prime}$–40$^{\prime}$–80$^{\prime}$ was employed to achieve an energy resolution of approximately 1.1~meV, ensuring that dynamic spin fluctuations at higher energies would not be misinterpreted as static magnetic correlations. A cylindrical-like single crystal of \Eu{}, with its long axis aligned along the crystallographic c-axis, was mounted on an aluminum holder. The crystal was oriented in the $(hk0)$ and $(0kl)$ scattering plane during the experiment but only data collected in $(hk0)$ scattering plane were used to perform magnetic structure refinements. The $(hk0)$ configuration ensures that neutron absorption majorly caused by europium can be treated as an approximately isotropic factor, which is effectively incorporated into the overall scale factor during refinement. The nuclear and magnetic structures were refined using the \textit{FullProf} suite.
\\

\clearpage

\subsection{Heat Capacity}

\begin{figure}[h]
    \centering
    \includegraphics[width=0.8\columnwidth]{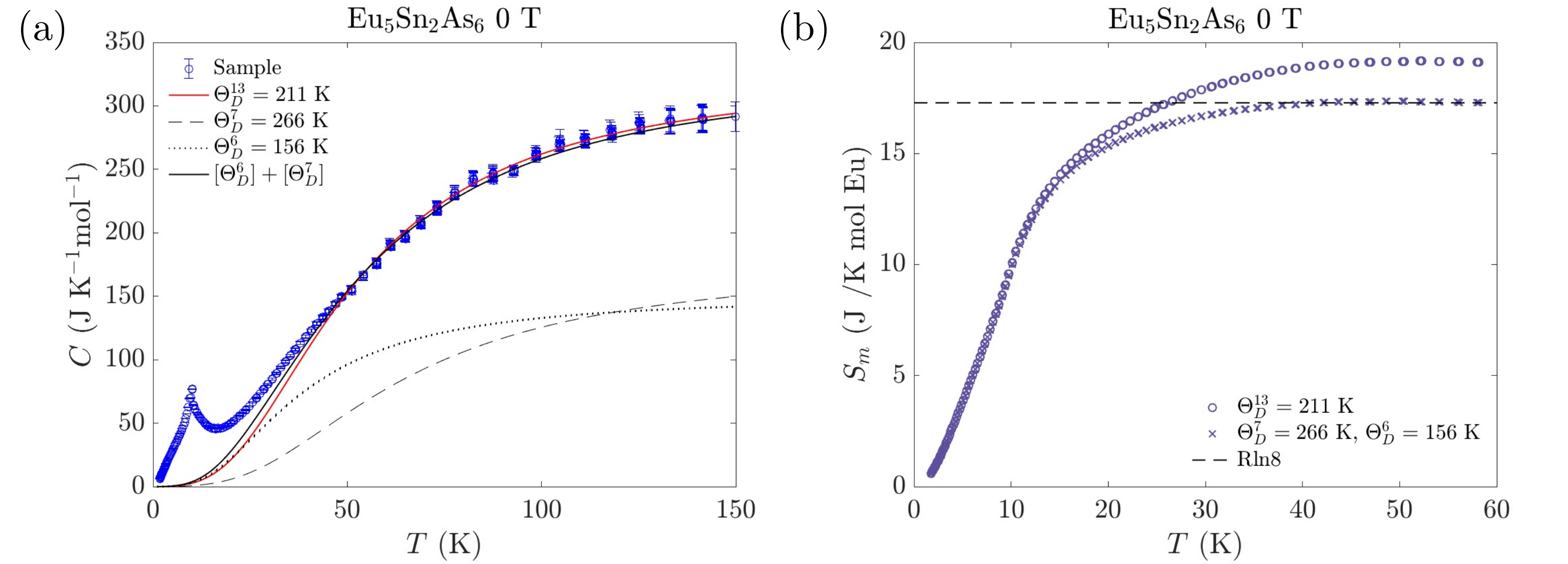}
    \caption{(a) Zero-field heat capacity of \Eu{}, including the two alternative Debye fits described in the main text. (b) Magnetic entropy in both cases.}
    \label{fig:SM_HC}
\end{figure}

Without a non-magnetic analogue the phonon contribution to the specific heat must be estimated. As shown in Fig.~\ref{fig:SM_HC}~(a), a Debye model with $\Theta_D=211$~K matches the data well at higher temperatures. However, as shown in Fig.~\ref{fig:SM_HC}~(b), in this case the released magnetic entropy reaches a value larger than the $R\ln{8}$ anticipated from the Eu$^{2+}$ $J=7/2$ moments. This also gives an anomalous non-monotonic tail in the magnetic heat capacity above $T_N$, hinting at the presence of a soft phonon mode. Instead, if we assume that the lightest atom (As) should have a lower Debye temperature and fit the data using a double Debye model then we can resolve both of these issues. The double Debye fitting shown in Fig.~\ref{fig:SM_HC}~(a) is that which has been subtracted from the total heat capacity to give the data shown in Fig.~\ref{fig:Fig2}~(a).
\\

The choice in dividing the Debye models is arbitrary, for example we can get similar results by instead assuming that the Sn atoms should have a lower Debye temperature, or fitting with more than two Debye models. However, we are always constrained by the fact that the magnetic subsystem should release $R\ln{8}$ of entropy. Regardless of the precise phonon contribution, the magnetic heat capacity remains finite to a value far larger than $T_N$. The observation that spin correlations persist to a temperature similar to that at which the colossal magnetoresistance and thermal magnetoresistance manifest therefore remains valid.
\\

\begin{figure}[h]
    \centering
    \includegraphics[width=0.8\columnwidth]{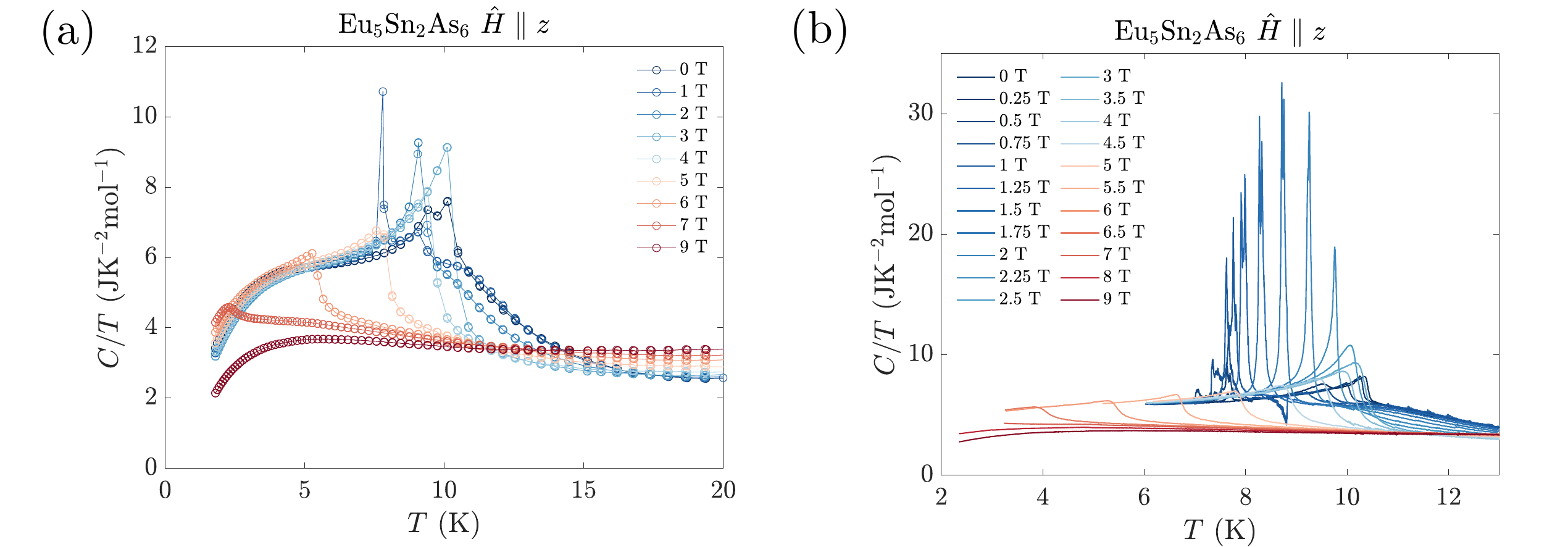}
    \caption{Heat capacity data used to produce the contour plot in Fig.~\ref{fig:Fig1}~(b). (a) shows data acquired using the relaxation time method and (b) using the slope method.}
    \label{fig:SM_HC_data}
\end{figure}

Fig.~\ref{fig:SM_HC_data} shows the heat capacity data used to produce the contour plot in Fig.~\ref{fig:Fig1}~(b), acquired through both the relaxation time and slope methods. The relaxation method necessarily averages over a small temperature window (in this case 2\% of the sample temperature) and will smooth over features sharper than this. The slope method was therefore used to measure the heat capacity more accurately through the phase transitions.

\clearpage

\subsection{Magnetisation}

\begin{figure}[h]
    \centering
    \includegraphics[width=0.8\columnwidth]{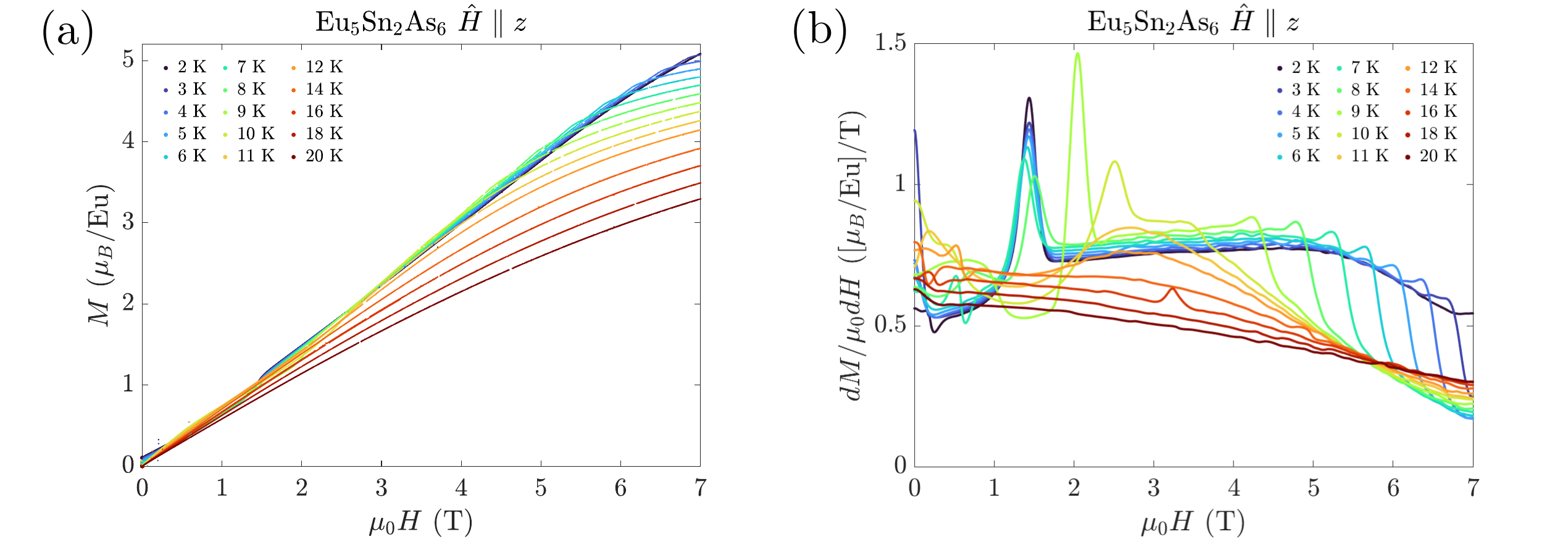}
    \caption{Magnetisation data used to produce the contour plot in Fig.~\ref{fig:Fig1}~(b). (a) shows the magnetisation and (b) shows the same data differentiated with respect to field.}
    \label{fig:SM_M_data}
\end{figure}

Fig.~\ref{fig:SM_M_data} shows the magnetisation data used to produce the contour plot in Fig.~\ref{fig:Fig1}~(b). The magnetisation differentiated with respect to field was calculated by fitting a smoothing spline function to the magnetisation data, then differentiating this function.

\clearpage

\subsection{Thermal Conductivity}

\begin{figure}[h]
    \centering
    \includegraphics[width=0.4\columnwidth]{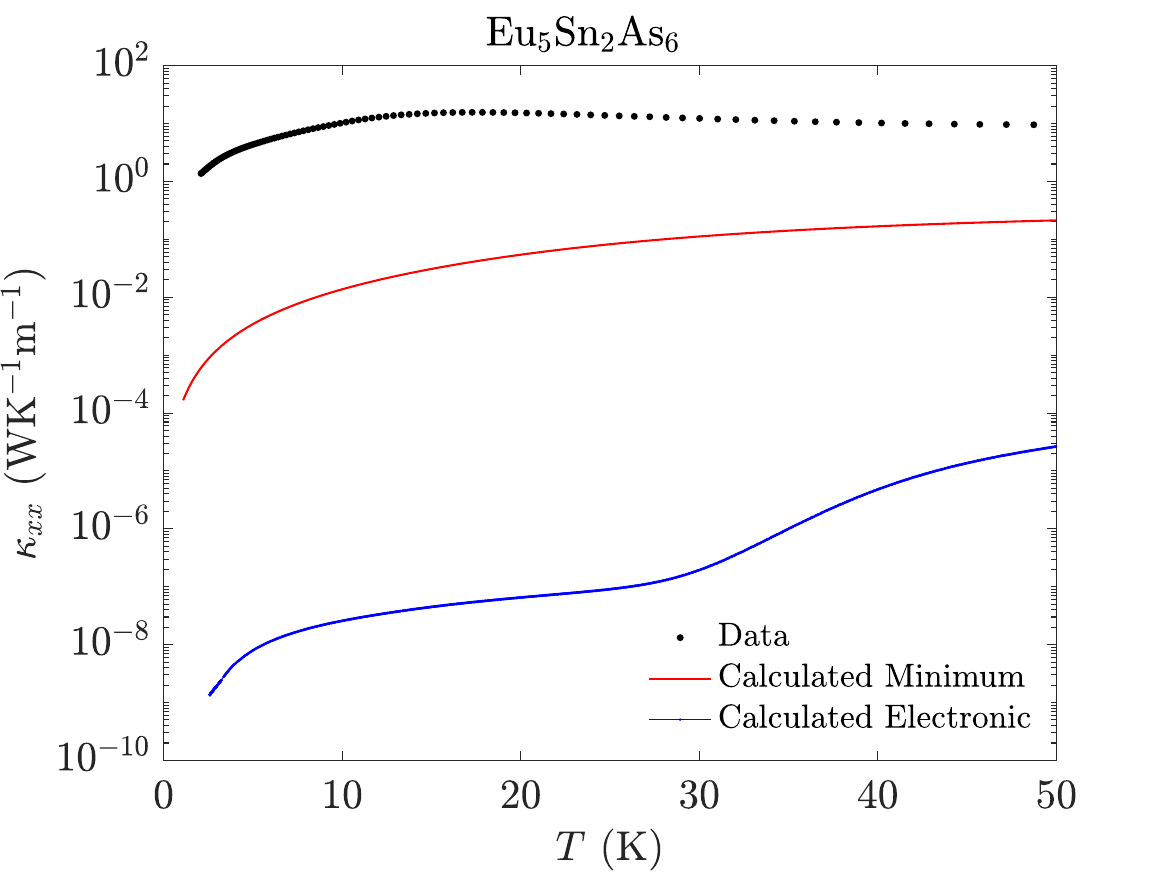}
    \caption{Zero-field thermal conductivity data of \Eu{} alongside the calculated electronic contribution and also the calculated minimum phonon conductivity, as described in the main text.}
    \label{fig:SM_Kxx_calcs}
\end{figure}

Fig.~\ref{fig:SM_Kxx_calcs} shows the zero-field thermal conductivity as compared to the calculated electronic and minimum phonon conductivities. The former is calculated by converting the electrical resistivity data in Fig.~\ref{fig:Fig2}~(b) to a thermal conductivity through the Wiedemann-Franz law. The latter is calculated by assuming a random walk between anharmonic oscillators, based on the procedure in \cite{Cahill1992}. 

\begin{figure}[h]
    \centering
    \includegraphics[width=0.8\columnwidth]{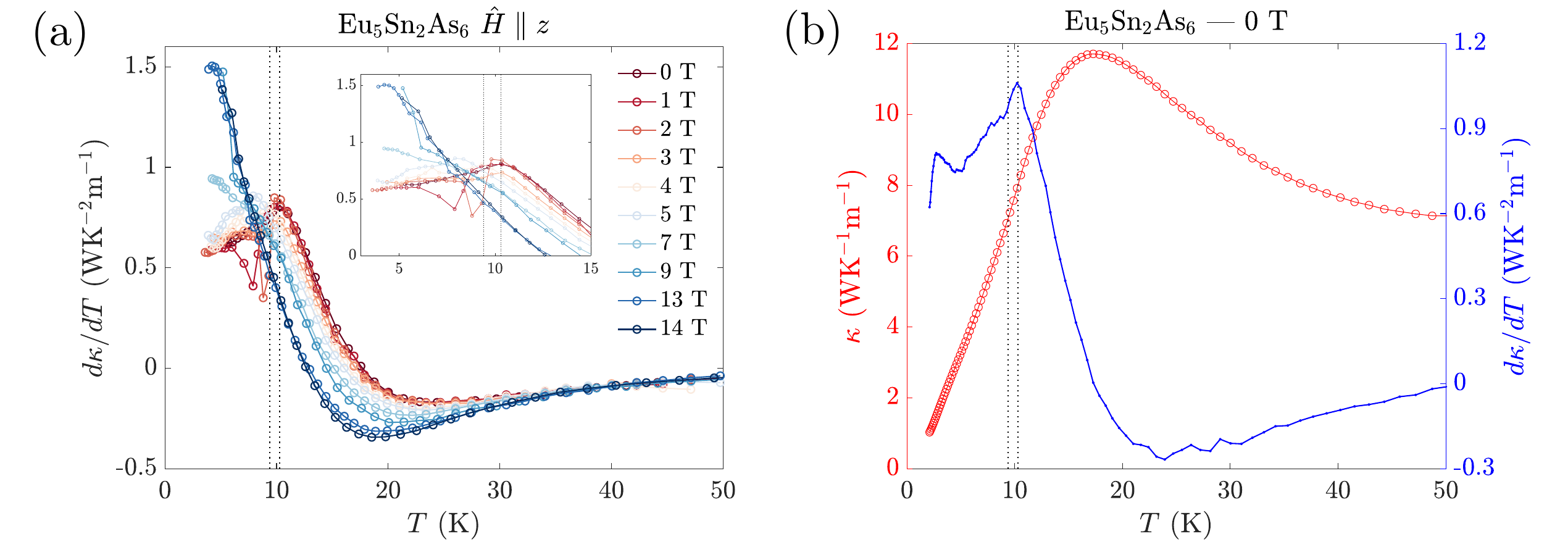}
    \caption{(a) Thermal conductivity data from Fig.~\ref{fig:Fig2}~(a) differentiated with respect to temperature, calculated using the finite difference method. (b) Zero-field thermal conductivity and differentiated thermal conductivity for a different sample of \Eu{}, with increased point density. In both cases the black dashed lines are at $T_N$ and $T_{N'}$ in zero-field.}
    \label{fig:SM_Kxx}
\end{figure}

Fig.~\ref{fig:SM_Kxx}~(a) shows the thermal conductivity from Fig.~\ref{fig:Fig2}~(a) differentiated with respect to temperature, in order to demonstrate the subtle impact of the magnetic phase transitions.

\clearpage

\subsection{Neutron Scattering}

Tables \ref{tab: BVs}, \ref{tab: selection} and \ref{tab:sym} show, respectively, the relevant irreducible representations of the little group \( G(\mathbf{k}=0) \) and their associated basis vectors, the selection rules for the magnetic Bragg peak positions for each, and the symmetry elements of the magnetic structure for each, as discussed in Appendix~A.

\begin{table}[h]
\begin{center}
\begin{tabular}{|c|c|c|c|c|c|c|} 
 \hline
 Atoms & Eu$_{1-1}$ & Eu$_{1-2}$ & Eu$_{2-1,2-2}$ & Eu$_{2-3,2-4}$ & Eu$_{3-1,3-2}$& Eu$_{3-3,3-4}$\\ \hline
 Position&(0,0,0)&(0.5,0.5,0)&($\pm$0.1230,$\pm$0.2421,0)&($\pm$0.3770,$\pm$0.7421,0)&($\pm$0.3708,$\pm$0.0668,0)&($\pm$0.1292,$\pm$0.5668,0)\\
 \hline
 $\Gamma_{1}$&(0,0,1)&(0,0,-1)&(0,0,1)&(0,0,-1)&(0,0,1)&(0,0,-1)\\\hline
 $\Gamma_{3}$&(0,0,1)&(0,0,1)&(0,0,1)&(0,0,1)&(0,0,1)&(0,0,1)\\\hline
 $\Gamma_{5}$,BV1&(1,0,0)&(-1,0,0)&(1,0,0)&(-1,0,0)&(1,0,0)&(-1,0,0)\\\hline
 $\Gamma_{5}$,BV2& (0,1,0)& (0,1,0)& (0,1,0)& (0,1,0)& (0,1,0)&(0,1,0)\\\hline
 $\Gamma_{7}$,BV1&(1,0,0)&(1,0,0)&(1,0,0)&(1,0,0)&(1,0,0)&(1,0,0)\\\hline
 $\Gamma_{7}$,BV2& (0,1,0)& (0,-1,0)& (0,1,0)& (0,-1,0)& (0,1,0)&(0,-1,0)\\\hline
\end{tabular}
\end{center}
 \caption{Atom positions and magnetic moments for different basis vectors.}
 \label{tab: BVs}
\end{table}

\begin{table}[h]
\begin{center}
\begin{tabular}{|c|c|c|c|c|} 
 \hline
 IR & (100) & (010) & (020) & (200) \\ 
 \hline
 $\Gamma_{1} $ & \checkmark & \checkmark & $\times$ & $\times$ \\ 
 \hline
 $\Gamma_{3}$ & $\times$ & $\times$ & \checkmark & \checkmark \\ 
 \hline
 $\Gamma_{5}$, BV1 & $\times$ & \checkmark & $\times$ & $\times$ \\ 
 \hline
 $\Gamma_{5}$, BV2 & $\times$ & $\times$ & $\times$ & \checkmark \\
 \hline
 $\Gamma_{7}$, BV1 & $\times$ & $\times$ & \checkmark & $\times$ \\ 
 \hline
 $\Gamma_{7}$, BV2 & \checkmark & $\times$ & $\times$ & $\times$ \\ 
 \hline
 $10$ K & \checkmark & $\times$ & \checkmark & $\times$ \\
 \hline
 $3.5$ K & \checkmark & \checkmark & \checkmark & \checkmark \\
 \hline
\end{tabular}
\end{center}
 \caption{Selection rules of Bragg peaks for each basis vector in different irreducible representation.}
 \label{tab: selection}
\end{table}

\begin{table}[h]
\begin{center}
    \begin{tabular}{|c|c|c|c|c|}
    \hline
        IR& $\Gamma_{7}$ & $\Gamma_{1}$ & $\Gamma_{3}$ & $\Gamma_{5}$ \\
    \hline
        $\{ 2_{001} | 0 \}$& $T$& 1& 1& $T$\\
    \hline
        $\{ 2_{010} | \tfrac{1}{2}\, \tfrac{1}{2}\, 0 \}$& $T$& 1& $T$& 1\\
    \hline
        $\{ 2_{100} | \tfrac{1}{2}\, \tfrac{1}{2}\, 0 \}$& 1& 1& $T$& $T$\\
    \hline
 $\{-1 | 0\}$& 1& 1& 1&1\\\hline
 $\{ m_{001} | 0 \}$& $T$& 1& 1&$T$\\\hline
 $\{ m_{010} | \frac{1}{2}\,\frac{1}{2}\, 0\}$& $T$& 1& $T$&1\\\hline
 $\{ m_{100} | \frac{1}{2}\,\frac{1}{2}\, 0\}$& 1& 1& $T$&$T$\\\hline
    \end{tabular}
\end{center}
    \caption{Non-trivial symmetry elements of magnetic structures determined by different irreducible representation, $'T'$ means addition time reversal symmetry operator is required.}
    \label{tab:sym}
\end{table}

\clearpage

\subsection{Electronic Structure Calculations}

Electronic structure calculations were carried out using density functional theory (DFT) as implemented in the Vienna Ab initio Simulation Package \cite{Kresse1996, Kresse1996_2}, using the PBE-GGA exchange correlation functional \cite{Perdew1996} and PAW pseudopotentials \cite{Kresse1999,Blochl1994}. The Brillouin zone was sampled with a $6 \times 18 \times 5$ Gamma-centered k-point mesh using the Bl\"{o}chl tetrahedron method \cite{Blochl1994_2}. A plane-wave-basis energy cut-off of 400~eV was used. The energy threshold for self-consistency was $10^{-6}$~eV, with self-consistency reached using the blocked-Davidson iteration scheme. The experimentally determined atomic structure was used---orthorhombic space group $Pbam$ (No. 55) with lattice parameters $a={12.3174}~\SI{}{\angstrom}$, $b={13.9827}~\SI{}{\angstrom}$ and $c={4.21344}~\SI{}{\angstrom}$.
\\

\begin{figure}[h]
    \centering
    \includegraphics[width=0.8\columnwidth]{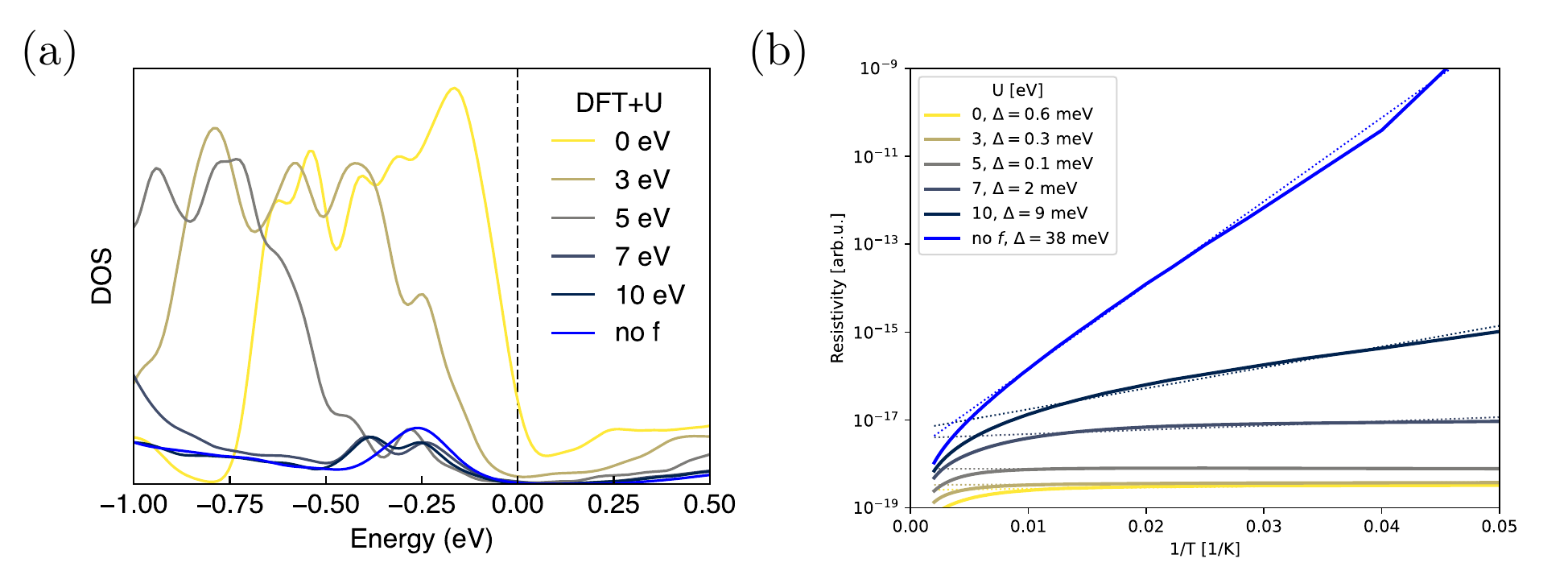}
    \caption{(a) Evolution of the total density of states with changes in the Hubbard $U$. (b) Resistivity calculated by the Boltzmann transport equation for different values of $U$. The dashed lines are fits to the Arrhenius model of thermally-activated carriers, $\rho \sim e^{\Delta/k_B T}$, fitted between 20 and 100~K.}
    \label{fig:SM_DFT}
\end{figure}

When included, the Eu $4f$-electrons were treated with DFT+U by the Dudarev method \cite{Dudarev1998}. For DFT+U, a range of $U$ values were tested. The primary effect of increasing $U$ was to shift the Eu $f$-band down and away from the Fermi level (Fig.~\ref{fig:SM_DFT}~(a)), although $U > 5$~eV also opened a small gap, reproducing the expected insulating behavior. Spin-orbit coupling was included, with the initial magnetic moments set to the experimental structure from neutron scattering; notably, the experimental magnetic configuration was preserved after DFT self-consistency, indicating it is at least a local minimum of the free energy function. We separately calculated the properties of \Eu{} in the paramagnetic regime by freezing the 7 Eu$^{2+}$ $4f$ electrons into the core in the pseudopotential, effectively ignoring the $f$ electrons.
\\

Electrical conductivity was calculated with the semiclassical Boltzmann transport equation as implemented in BoltzTraP2 \cite{Madsen2018}. Constant relaxation time was assumed. The Brillouin zone sampling was enhanced by a factor of 150 for the calculation of transport properties. As discussed in the main text, the paramagnetic (no $f$-electron) case matches the experimental data, with a gap of $\Delta = 38$~meV (see Fig.~\ref{fig:SM_DFT}~(b)).

\end{document}